\documentclass[11pt]{article}
\usepackage[margin=1in]{geometry}
\usepackage{amsmath,amssymb,amsthm,booktabs}
\usepackage[authoryear,round]{natbib}
\usepackage[hidelinks]{hyperref}
\usepackage{graphicx}
\usepackage[font=small]{caption}
\usepackage{listings}
\usepackage{quantikz}
\theoremstyle{definition}
\newtheorem{theorem}{Theorem}
\newtheorem{proposition}{Proposition}
\newtheorem{corollary}{Corollary}

\newtheorem{remark}{Remark}
\newcommand{\E}{\mathbb E}
\newcommand{\ii}{\mathrm i}
\newcommand{\qcwtket}[1]{\lvert #1\rangle}
\newcommand{\qcwtbra}[1]{\langle #1\rvert}
\title{A Quantum Continuous Wavelet Transform}
\author{Brani Vidakovic\\
\small Department of Statistics, Texas A\&M University}
\date{}
\begin{document}
\maketitle

\begin{abstract}
We construct a quantum continuous wavelet transform (QCWT) on a finite scale--translation grid whose scales may be non-dyadic. A quantum Fourier transform places the amplitude-encoded signal in the frequency basis, the scale-dependent multiplier is block-encoded with one flag qubit, and an inverse Fourier transform returns the complex, quadrature-weighted CWT coefficients in the success branch. For heralded circuits that act once on one copy of the input with an input-independent output factor, we bound the success probability and attain the bound by a frequency-conditioned scale preparation. When the scale prior is prepared independently of frequency, normalizing the multiplier separately at each scale is optimal and raises the success probability by factors of 4.9 and 14.6 in our examples; on logarithmic grids this architecture loses, relative to the bound, a factor of about the admissibility constant over the squared spectral peak times the logarithmic scale range. We give a fixed-point resource estimate for a reversible Mexican-hat multiplier, the multinomial law of scale--translation measurements, and Hadamard tests for complex coefficients. We show that a Fourier-basis measurement of the input, followed by classical randomization, reproduces the flag and scale outcomes of every such heralded circuit exactly, so it is at least as informative for global scale energies; a simulation study of Hurst exponent estimation agrees. What the QCWT adds is the joint scale--translation law: two chirps with identical Fourier-basis and time-basis measurement laws have QCWT laws at total-variation distance 0.27.
\end{abstract}


\section{Introduction}

The continuous wavelet transform (CWT) analyzes a signal through translated and dilated copies of a mother wavelet \citep{GrossmannMorlet1984,Daubechies1992,Mallat1999}. Because the scale may vary continuously, the CWT is the natural representation for scale ridges, local regularity, and power laws over selected scale ranges, and its squared modulus, the scalogram, is a standard tool in geophysics, turbulence, and the analysis of self-similar processes \citep{TorrenceCompo1998,Flandrin1992,AbryVeitch1998}. On a digital computer one evaluates the CWT on a finite grid. A quantum implementation inherits the discretization problem and adds a constraint of its own: a wavelet filter changes both magnitudes and phases in the frequency domain, whereas quantum evolution must preserve the norm of the full state.

The $\varepsilon$-decimated quantum nondecimated wavelet transform (QNDWT) of \citet{Vidakovic2025} starts from an orthogonal discrete wavelet unitary and varies its circular alignment through a shift register. It yields nondecimated detail coefficients on a fixed dyadic family of levels, and because every operation is unitary and the decimated transforms are orthogonal, it requires no postselection. The present paper takes the opposite route. It discretizes the CWT of a chosen mother wavelet directly, giving what we call a quantum continuous wavelet transform (QCWT), so each prescribed scale has its own Fourier multiplier, the grid may contain any number of intermediate scales between dyadic octaves, and no orthonormal filter bank is required. The price of this flexibility is that a scale-dependent multiplier is not unitary, so it must be embedded in a larger unitary by block encoding \citep{ChildsWiebe2012,Gilyen2019}, and the CWT appears only in a heralded branch. Linear combinations of unitaries also underlie the construction of \citet{BagherimehrabAspuruGuzik2024}, which implements multilevel discrete and wavelet-packet transforms of any order; there the target is an orthogonal transform, so amplitude amplification with a known success probability makes it deterministic. The sampled CWT is a redundant, non-unitary frame map, so its success probability depends on the input and on the scale grid. How large that branch is, how its probability depends on the scale grid, and what statistical information a measurement of it carries are the central questions of the paper.

Two caveats frame the discussion. Preparing the amplitude-encoded state of an arbitrary vector of length $N$ requires $\Theta(N)$ elementary gates in general \citep{PleschBrukner2011,ZhangLiYuan2022}, so a quantum CWT cannot accelerate the classical CWT of data already held in classical memory, for which a fast Fourier transform per scale is the appropriate baseline \citep{Aaronson2015}. Its use is as a subroutine on states produced coherently by an earlier quantum process or an efficient structured loader \citep{GroverRudolph2002}. Moreover, a measurement returns samples from a scalogram law, not the coefficient array, so statistical procedures must be designed for that law. We therefore state costs in two currencies, the gate count of the circuit acting on a given state and the number of state preparations needed for a given statistical precision.

The contributions are as follows. Section~\ref{sec:model} records the sampled periodic model of the CWT, including a mirror extension for nonperiodic records that costs one qubit and a layer of controlled NOT gates. Section~\ref{sec:circuit} constructs the flagged circuit, proves an upper bound on the success probability of any heralded circuit producing the CWT coefficient state, attained by a frequency-conditioned scale preparation, and proves that scale-dependent normalization is optimal within the simpler architecture whose scale prior does not depend on frequency. We quantify the gap between the two architectures on logarithmic grids, give a fixed-point resource estimate with error propagation for a reversible Mexican-hat multiplier, and show that normalization by the spectral envelope gives an exactly uniform scale prior at negligible loss. Section~\ref{sec:readout} develops the multinomial measurement model, estimators of unweighted scale energies, Hadamard tests for individual complex coefficients, a Fourier-basis measurement that reproduces the QCWT flag and scale outcomes exactly, and the relation to the QNDWT. Section~\ref{sec:examples} verifies the construction on a two-tone signal, on a 48-scale analysis of the Doppler signal, on two wavelets with complex-valued multipliers or coefficients, and on a pair of chirps that the Fourier-basis and time-basis measurements cannot distinguish but the QCWT can. Section~\ref{sec:hurst} studies Hurst exponent estimation from fractional Brownian paths, separating the effect of a calibrated mean curve from that of covariance weighting and comparing the quantum CWT with the Fourier baseline at equal numbers of input preparations. Section~\ref{sec:discussion} discusses resources and open problems.

\section{The Continuous Transform and Its Sampled Periodic Model}
\label{sec:model}

\subsection{Continuous Conventions}

Let $a>0$, $b\in\mathbb R$, and $\psi_{a,b}(t)=a^{-1/2}\psi((t-b)/a)$. With $\widehat f(\omega)=\int f(t)e^{-\ii\omega t}\,dt$ and its inverse carrying the factor $1/(2\pi)$, the CWT and its Fourier representation are
\begin{eqnarray}
W_f(a,b)&=&\int_{\mathbb R} f(t)\overline{\psi_{a,b}(t)}\,dt \nonumber\\
&=&\frac{1}{2\pi}\int_{\mathbb R}\widehat f(\omega)\sqrt a\,\overline{\widehat\psi(a\omega)}e^{\ii\omega b}\,d\omega . \label{eq:continuous}
\end{eqnarray}
Equivalently, $W_f(a,\cdot)=f*\widetilde\psi_a$ with $\widetilde\psi_a(t)=a^{-1/2}\overline{\psi(-t/a)}$. The factor $\sqrt a$ in the Fourier multiplier corresponds to the $L^2$ normalization of each wavelet atom.

For positive scales the natural energy measure is $da\,db/a^2$ \citep{Daubechies1992}. Set $C_\pm=\int_0^\infty |\widehat\psi(\pm u)|^2\,du/u$. Plancherel's identity and the substitution $u=a|\omega|$ give
\begin{eqnarray}
\int_0^\infty\int_{\mathbb R}|W_f(a,b)|^2\,db\,\frac{da}{a^2}
 &=& \frac{1}{2\pi}\int_{\mathbb R}|\widehat f(\omega)|^2 C_{\operatorname{sgn}(\omega)}\,d\omega . \label{eq:plancherel}
\end{eqnarray}
For a nonzero real admissible wavelet the two constants coincide, $C_+=C_-=C_\psi$, and the scaled CWT is an isometry into a redundant coefficient space. For a one-sided analytic wavelet the identity holds with its nonzero constant on the matching frequency half-line. A finite scale range and a finite quadrature generally lose exact tightness, and Section~\ref{sec:circuit} shows that the loss controls the success probability of the quantum circuit.

\subsection{Independent Discretization of Scale and Translation}

Let $N=2^n$, let $T=N\Delta t$ be the period, and take $b_j=j\Delta t$ for $j=0,\ldots,N-1$. Let $x$ contain the samples of a $T$-periodic trigonometric polynomial on the $N$ signed DFT frequencies; equivalently, start from $x$ and define its periodic trigonometric interpolant on those frequencies. Denote by $F$ the unitary DFT with entries $F_{kt}=N^{-1/2}\exp(-2\pi\ii kt/N)$, and let $\omega_k=2\pi\widetilde k/T$, where $\widetilde k$ is the signed frequency index in DFT order. We evaluate the periodic CWT at prescribed positive scales $a_s$, $s=0,\ldots,M-1$, using the periodized wavelet. Applying (\ref{eq:continuous}) to each Fourier mode gives
\begin{eqnarray}
m_{sk}&=&\sqrt{a_s}\,\overline{\widehat\psi(a_s\omega_k)},\qquad D_s=\operatorname{diag}_{k}(m_{sk}), \nonumber\\
c_s &=& F^\dagger D_s F x,\qquad c_{sj}=W_x(a_s,b_j). \label{eq:discretecwt}
\end{eqnarray}
Here $W_x$ denotes the transform of the interpolant, and the normalization of $F$ cancels between the forward and inverse transforms. For real data at even $N$ one can remove the Nyquist bin or fix its interpolation convention explicitly.

The scales may be arbitrary or placed on a dense logarithmic grid. Positive quadrature weights $w_s$ approximate the measure $da/a^2$. With $u=\log a$ this measure becomes $e^{-u}du$, so midpoint weights on a logarithmic grid are $w_s\approx \Delta u_s/a_s$; for $V$ voices per octave, $\Delta u_s=\ln 2/V$. The weights determine the sampling distribution and may be chosen separately from the scale locations. For a periodic signal with absolutely summable Fourier coefficients $f_k$, replacing the signal by its Fourier truncation to an index set $\mathcal K$ changes $W_f(a,b)$ by at most
$$
\sqrt{a_{\max}}\,\|\widehat\psi\|_\infty\sum_{k\notin\mathcal K}|f_k|
$$
uniformly in $b$ for $a\le a_{\max}$. For samples of a non-bandlimited signal, aliasing in the interpolant must also be accounted for.

A nonperiodic record requires boundary treatment, because the periodic model joins the end of the record to its beginning and the resulting jump contaminates coarse scales. The even (mirror) extension $(x_0,\ldots,x_{N-1},x_{N-1},\ldots,x_0)$ of length $2N$ is continuous at both junctions, so a wavelet with at least two vanishing moments sees only a kink there. On a quantum computer the extension costs one qubit and $n$ controlled NOT gates: a Hadamard gate on a new most significant qubit followed by NOT gates on all $n$ data qubits controlled by it maps $\qcwtket{x}$ to the normalized amplitude encoding of the extended vector, because bitwise complementation maps the index $t$ to $N-1-t$. Alternatively, one may pad and restrict the translations to the interior, at the cost of a translation-dependent exposure that must be accounted for in estimation.

\section{Flagged Block Encoding of the CWT}
\label{sec:circuit}

\subsection{Construction}

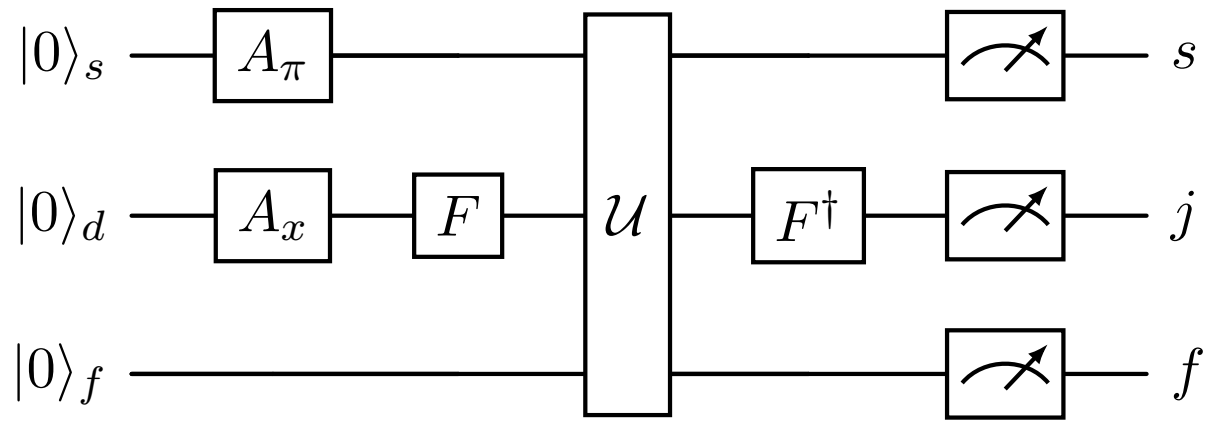
\begin{figure}[htbp]
\centering
\begin{quantikz}[column sep=0.55cm,row sep=0.45cm]
 \lstick{$\qcwtket{0}_s$} & \gate{A_\pi} & \qw      & \gate[3]{\mathcal U} & \qw                & \meter{} & \rstick{$s$} \\
 \lstick{$\qcwtket{0}_d$} & \gate{A_x}   & \gate{F} &                     & \gate{F^\dagger}   & \meter{} & \rstick{$j$} \\
 \lstick{$\qcwtket{0}_f$} & \qw          & \qw      &                     & \qw                & \meter{} & \rstick{$f$}
\end{quantikz}
\caption{Quantum CWT on a finite scale--translation grid. The joint gate $\mathcal U$ uses the scale label $s$ and the frequency label $k$ to select $U_{sk}$ on the flag. Retaining outcomes with $f=0$ gives the quadrature-weighted scalogram law (\ref{eq:law}).}
\label{fig:qcwt-circuit}
\end{figure}

Suppose $\qcwtket{x}=\sum_t x_t\qcwtket t/\|x\|_2$ can be prepared by a unitary $A_x$. For each scale choose a normalization $\alpha_s\ge\max_k|m_{sk}|$, and define the scale prior
$$
\pi_s=\frac{w_s\alpha_s^2}{\sum_r w_r\alpha_r^2}.
$$
A unitary $A_\pi$ prepares $\sum_s\sqrt{\pi_s}\qcwtket s$ on a scale register of $\lceil\log_2M\rceil$ qubits, leaving unused labels unoccupied. Apply $F$ to the data register, and put
$$
d_{sk}=m_{sk}/\alpha_s,\qquad r_{sk}=\sqrt{1-|d_{sk}|^2},\qquad
U_{sk}=\begin{pmatrix}d_{sk}&-r_{sk}\\ r_{sk}&\overline{d_{sk}}\end{pmatrix}.
$$
Each block $U_{sk}$ is unitary and its top-left entry carries the complete complex multiplier. The operator $\mathcal U=\sum_{s,k}\qcwtket{s,k}\qcwtbra{s,k}\otimes U_{sk}$ acts on a flag qubit initialized to $\qcwtket0$, and $F^\dagger$ returns the data register to the translation basis. The circuit is shown in Fig.~\ref{fig:qcwt-circuit}. The original formulation with a common normalization $\alpha=\max_{s,k}|m_{sk}|$ is the special case $\alpha_s\equiv\alpha$, for which $\pi_s\propto w_s$.

\begin{proposition}[CWT in the success branch]\label{prop:success}
With the flag ordered last, the unnormalized $\qcwtket0$ branch of the final state is
$$
\frac{1}{\|x\|_2\big(\sum_r w_r\alpha_r^2\big)^{1/2}}\sum_{s=0}^{M-1}\sum_{j=0}^{N-1}\sqrt{w_s}\,c_{sj}\qcwtket{s,j,0}.
$$
The success probability is
\begin{eqnarray}
P_0=\frac{\sum_s w_s\|c_s\|_2^2}{\|x\|_2^2\sum_r w_r\alpha_r^2},
\label{eq:P0}
\end{eqnarray}
and, when $P_0>0$, the conditional law of a scale--translation measurement is
\begin{eqnarray}
\Pr(s,j\mid 0)&=&\frac{w_s|c_{sj}|^2}{\sum_{r,\ell}w_r|c_{r\ell}|^2},
\label{eq:law}
\end{eqnarray}
which does not depend on the choice of normalizations $\alpha_s$.
\end{proposition}

\begin{proof}
Immediately before $\mathcal U$ the amplitude at $(s,k,0)$ is $\sqrt{\pi_s}(Fx)_k/\|x\|_2$. The $\qcwtket0$ component after $\mathcal U$ multiplies it by $m_{sk}/\alpha_s$, and $\sqrt{\pi_s}/\alpha_s=\sqrt{w_s}/(\sum_rw_r\alpha_r^2)^{1/2}$. Applying $F^\dagger$ within each scale gives (\ref{eq:discretecwt}) with the stated factor. Squared norms give the success probability and the conditional law.
\end{proof}

The translation weight $\Delta t$ cancels in (\ref{eq:law}), and the complex coefficients are present as amplitudes in the conditional state. The compensation between $\alpha_s$ and $\pi_s$ is the whole point: a larger prior mass at a scale whose multiplier must be divided by a larger constant leaves the joint amplitude unchanged up to a global factor.

\begin{remark}[Gate realization]
Writing $d=|d|e^{\ii\phi}$, the block $U_{sk}$ equals $P_\phi R_y(2\arccos|d|)P_\phi$ with $P_\phi=\operatorname{diag}(e^{\ii\phi/2},e^{-\ii\phi/2})$, where $\phi=0$ when $d=0$. Both the angle and the phase depend on $(s,k)$. For the Mexican hat and for the analytic Morlet wavelet the multiplier is real and nonnegative, so the phase gates are absent and $\mathcal U$ is a multiplexed $Y$ rotation; complex coefficients nevertheless arise for the Morlet wavelet because its multiplier vanishes on negative frequencies. For an odd real wavelet such as the first derivative of a Gaussian the multiplier is purely imaginary, so the phase gates are required while the coefficients are real; Section~\ref{sec:complex} verifies both cases.
\end{remark}

\subsection{Optimal Normalization and the Success Probability}

Let $q_k=\sum_s w_s|m_{sk}|^2$ be the frequency response of the weighted scale grid and $q_{\max}=\max_kq_k$. Unitarity of $F$ gives the finite-frame identity
\begin{eqnarray}
\sum_s w_s\|c_s\|_2^2 &=& \sum_k q_k |(Fx)_k|^2 , \label{eq:frame}
\end{eqnarray}
so the weighted coefficient map is a frame on any band $\mathcal B$ on which $0<A\le q_k\le B$, and (\ref{eq:P0}) becomes $P_0=\sum_kq_k|(Fx)_k|^2/(\|x\|_2^2\sum_rw_r\alpha_r^2)$. Two optimality questions arise, one about any circuit that produces the CWT coefficient state and one about the architecture of Fig.~\ref{fig:qcwt-circuit}. We separate them.

\begin{theorem}[Bound for single-copy heralded constructions]\label{thm:global}
Consider any circuit that applies a fixed unitary, not depending on $x$, once to a single copy of $\qcwtket x$ and ancillas in fixed initial states and heralds success by a fixed outcome of some ancillas, and suppose that for every input $x$ the unnormalized heralded state is $\gamma\sum_{s,j}\sqrt{w_s}\,c_{sj}\qcwtket{s,j}/\|x\|_2$ with a constant $\gamma$ independent of $x$. Then $|\gamma|^2\le1/q_{\max}$, and therefore
\begin{eqnarray}
P_0\le\frac{\sum_kq_k|(Fx)_k|^2}{q_{\max}\|x\|_2^2}.
\label{eq:globalbound}
\end{eqnarray}
The bound is attained by the following circuit: after $F$, rotate the flag so that its $\qcwtket0$ amplitude is $(q_k/q_{\max})^{1/2}$, controlled by the frequency label $k$; on the flag-zero branch prepare the scale state $\sum_s\sqrt{w_s}\,m_{sk}q_k^{-1/2}\qcwtket s$, controlled by $k$; apply $F^\dagger$.
\end{theorem}

\begin{proof}
The heralded branch is $K\qcwtket{x}$ for a linear map $K$ of operator norm at most one, a block of a unitary. The required map is $\gamma C$ with $C x=\sum_{s,j}\sqrt{w_s}c_{sj}\qcwtket{s,j}$, and $C^\dagger C=F^\dagger\operatorname{diag}(q_k)F$ by (\ref{eq:discretecwt}), so $\|C\|^2=q_{\max}$ and $|\gamma|^2q_{\max}\le1$. The success probability is $|\gamma|^2\|Cx\|^2/\|x\|_2^2$, which gives (\ref{eq:globalbound}) by (\ref{eq:frame}). In the stated circuit the scale state is normalized because $\sum_sw_s|m_{sk}|^2=q_k$, and the amplitude at $(s,k,0)$ is $\sqrt{w_s}m_{sk}(Fx)_k/(\sqrt{q_{\max}}\|x\|_2)$, so $|\gamma|^2=1/q_{\max}$.
\end{proof}

For an input whose spectrum lies where $q_k$ is close to $q_{\max}$ the bound is close to one. In our examples it is $0.986$ for the two-tone signal and $0.783$ for the Doppler signal. The hypotheses exclude amplitude amplification (Section~\ref{sec:aa}), which uses the preparation $A_x$ and its inverse repeatedly to reflect about $\qcwtket x$, so the amplified circuit is neither a single use of one copy nor a fixed unitary, and its success probability may exceed (\ref{eq:globalbound}). The cost of the frequency-conditioned construction lies in the controlled preparation of an $M$-dimensional scale state for each frequency, which in general is a multiplexer of size $MN$; on a logarithmic grid the conditional scale amplitudes are a fixed profile $\widehat\psi(2^{v})$ evaluated at $v=\log_2a_0+s/V+\log_2|\omega_k|$, a translate in $s$ by a non-integer amount, and whether this structure yields an efficient preparation is open. We therefore treat (\ref{eq:globalbound}) as the benchmark and analyze the architecture of Fig.~\ref{fig:qcwt-circuit}, in which the scale prior is prepared once, independently of frequency.

\begin{theorem}[Optimal normalization for an independent scale prior]\label{thm:optimal}
Consider any circuit that prepares $\sum_s\sqrt{\pi_s}\qcwtket s\otimes F\qcwtket x$ with $\pi$ independent of $x$ and of $k$, applies on the flag a unitary $U_{sk}$ controlled by $(s,k)$ with top-left entry $d_{sk}$, applies $F^\dagger$, and whose flag-zero branch equals $\gamma\sum_{s,j}\sqrt{w_s}c_{sj}\qcwtket{s,j,0}/\|x\|_2$ for every input $x$. Then
$$
|\gamma|^2\le\frac{1}{\sum_s w_s\alpha_s^{\ast2}},\qquad \alpha_s^\ast=\max_k|m_{sk}|,
$$
with equality for the construction of Proposition~\ref{prop:success} with $\alpha_s=\alpha_s^\ast$. For every input the success probability of this architecture is therefore the bound (\ref{eq:globalbound}) multiplied by
\begin{eqnarray}
\rho=\frac{q_{\max}}{\sum_sw_s\alpha_s^{\ast2}}\le1,
\label{eq:rho}
\end{eqnarray}
and the common normalization $\alpha=\max_s\alpha_s^\ast$ multiplies it further by $\sum_sw_s\alpha_s^{\ast2}/(\alpha^2\sum_sw_s)\le1$.
\end{theorem}

\begin{proof}
Before $F^\dagger$ the flag-zero amplitude at $(s,k)$ is $\sqrt{\pi_s}\,d_{sk}(Fx)_k/\|x\|_2$, and the requirement on the output, after applying $F$ within each scale, is $\sqrt{\pi_s}\,d_{sk}=\gamma\sqrt{w_s}\,m_{sk}$ for every $(s,k)$, since inputs with a single nonzero Fourier coefficient are admissible. Because $|d_{sk}|\le1$, this forces $\pi_s\ge|\gamma|^2w_s\alpha_s^{\ast2}$, and summing over $s$ gives the bound. The success probability is $|\gamma|^2\sum_kq_k|(Fx)_k|^2/\|x\|_2^2$, and the remaining statements follow by comparison with (\ref{eq:globalbound}).
\end{proof}

The factor $\rho$ is the price of a frequency-independent prior, not a fundamental cost of producing a CWT state, and it is the same for every input. It has a transparent form on logarithmic grids. A frequency in the interior of the covered band is seen by every scale through $|\widehat\psi(a_s\omega)|^2$, and the Riemann sum $q_k=\sum_sw_sa_s|\widehat\psi(a_s\omega_k)|^2$ with $w_s=\Delta u/a_s$ approximates $\int|\widehat\psi(a\omega)|^2da/a=C_\psi$, while $\sum_sw_s\alpha_s^{\ast2}\approx\|\widehat\psi\|_\infty^2\ln(a_{\max}/a_{\min})$.

\begin{corollary}\label{cor:log}
Consider a logarithmic grid $a_s=a_{\min}2^{s/V}$ spanning $O$ octaves with weights $w_s=\ln2/(Va_s)$, and suppose that (i) the spectral peak of every scale is resolved by the frequency grid and lies below the Nyquist frequency, so that $\alpha_s^\ast\approx\sqrt{a_s}\|\widehat\psi\|_\infty$, which requires $a_{\min}$ bounded below and $a_{\max}$ small relative to $N$, and (ii) the scale interval is wide enough that, for some interior frequency, the covered scales capture most of the admissibility integral. Then
$$
\rho\approx\frac{C_\psi}{\|\widehat\psi\|_\infty^2\,O\ln 2},
$$
independently of $V$. With the common normalization the corresponding factor is approximately $(C_\psi/\|\widehat\psi\|_\infty^2)\{V(2^{1/V}-1)/\ln2\}2^{-O}$, which decays exponentially in the number of octaves.
\end{corollary}

\begin{figure}[htbp]
\centering
\includegraphics[width=0.55\linewidth]{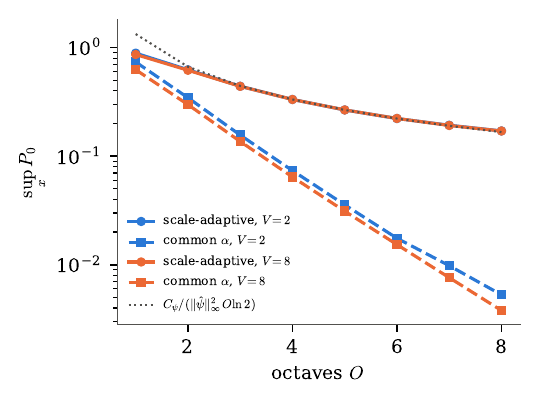}
\caption{Maximal success probability $\sup_xP_0=\rho$ of the independent-prior Mexican-hat QCWT on logarithmic grids starting at one sample spacing, $N=1024$, as a function of the number of octaves $O$, for $V=2$ and $V=8$ voices per octave, with scale-adaptive (circles) and common (squares) normalization. The dotted line is the approximation of Corollary~\ref{cor:log}, which is accurate for $3\le O\le8$ at this $N$. The frequency-conditioned construction of Theorem~\ref{thm:global} has supremum one.}
\label{fig:success}
\end{figure}

The asymptotic statement lets $O$ grow with $a_{\min}$ fixed and with $N$ large enough that condition (i) holds at the largest scale. For the Mexican hat, $\widehat\psi(u)=u^2e^{-u^2/2}$, one has $C_\psi=\int_0^\infty u^3e^{-u^2}du=1/2$ and $\|\widehat\psi\|_\infty=2/e$, attained at $u=\sqrt2$, so $\rho\approx1.33/O$. Table~\ref{tab:regime} and Fig.~\ref{fig:success} give exact finite-grid values. Condition (ii) fails for $O\le2$, where the approximation exceeds the exact value and even exceeds one. Condition (i) begins to fail at $N=1024$ for $O=9$, where the largest scale is 256 samples and its spectral peak at $\omega=\sqrt2/256$ is resolved by only about two Fourier bins; there the exact value $0.161$ exceeds the approximation $0.148$, and at $N=4096$ the agreement is restored. For $3\le O\le8$ the approximation is accurate to the third decimal at both $N$.

\begin{table}[htbp]
\centering
\caption{Exact finite-grid values of $\rho$ in (\ref{eq:rho}) for the Mexican hat, which equal $\sup_xP_0$ for the independent-prior architecture, on logarithmic grids beginning at one sample spacing. Exact: $\alpha_s=\alpha_s^\ast$; envelope: $\alpha_s=\sqrt{a_s}\|\widehat\psi\|_\infty$; common: $\alpha=\max_s\alpha_s^\ast$. The last column is the approximation of Corollary~\ref{cor:log}. The global bound (\ref{eq:globalbound}) has supremum one in every case.}
\label{tab:regime}
{\small
\begin{tabular}{c|cccc|c|c}
\toprule
 & \multicolumn{4}{c|}{Scale-adaptive} & Common & \\
$O$ & $N\!=\!1024$, $V\!=\!2$ & $N\!=\!1024$, $V\!=\!8$ & envelope, $V\!=\!8$ & $N\!=\!4096$, $V\!=\!8$ & $N\!=\!1024$, $V\!=\!8$ & $1.33/O$\\
\midrule
1 & 0.889 & 0.864 & 0.864 & 0.864 & 0.6255 & 1.333 \\
2 & 0.625 & 0.619 & 0.619 & 0.619 & 0.2986 & 0.666 \\
3 & 0.441 & 0.440 & 0.440 & 0.440 & 0.1366 & 0.444 \\
4 & 0.333 & 0.333 & 0.333 & 0.333 & 0.0643 & 0.333 \\
5 & 0.267 & 0.267 & 0.266 & 0.266 & 0.0312 & 0.267 \\
6 & 0.222 & 0.223 & 0.222 & 0.222 & 0.0153 & 0.222 \\
7 & 0.193 & 0.192 & 0.190 & 0.190 & 0.0076 & 0.190 \\
8 & 0.171 & 0.171 & 0.167 & 0.167 & 0.0038 & 0.167 \\
9 & 0.158 & 0.161 & 0.148 & 0.149 & 0.0036 & 0.148 \\
\bottomrule
\end{tabular}
}
\end{table}

The exact optimum $\alpha_s=\alpha_s^\ast$ generally gives a nonuniform prior; in the Doppler example of Section~\ref{sec:examples} it ranges from $0.0186$ to $0.0210$. Normalizing instead by the spectral envelope, $\alpha_s=\sqrt{a_s}\|\widehat\psi\|_\infty$, satisfies $\alpha_s\ge\max_k|m_{sk}|$ for any bounded wavelet spectrum and, with $w_s=\Delta u/a_s$, gives the prior $\pi_s\propto\Delta u_s$, exactly uniform on a uniform logarithmic grid. The loss is negligible in our examples: $P_0=0.3161$ instead of $0.3171$ for the two-tone signal and $0.1740$ instead of $0.1755$ for the Doppler signal. A uniform state on $M$ labels is a Hadamard layer when $M$ is a power of two; for other $M$, such as the 48 scales of the Doppler example, it requires a uniform-superposition preparation on $\lceil\log_2M\rceil$ qubits with $O(\log M)$ gates \citep{ShuklaVedula2024}, or one may choose $M$ to be a power of two. The common normalization instead requires the nonuniform prior $\pi_s\propto1/a_s$, which in the Doppler example ranges over a factor of 59.

\subsection{Multiplier Synthesis and Error Propagation}\label{sec:synth}

The gate $\mathcal U$ dominates the cost. A table-driven multiplexer of $MN$ rotations, decomposed into uniformly controlled rotations, uses $O(MN)$ controlled NOT gates and $MN$ arbitrary-angle rotations \citep{Mottonen2004}, $24{,}576$ rotations for the Doppler example; this serves as a reference. With the envelope normalization the Mexican-hat multiplier depends on $(s,k)$ only through $u=a_s|\omega_k|$,
$$
d_{sk}=g(u)=e\,t\,e^{-t},\qquad t=u^2/2,
$$
and the flag rotation angle is $\theta=2\arccos g$. A reversible circuit computes $|\widetilde k|$ from the signed frequency label, looks up the constant $a_s\cdot2\pi/N$ from a table indexed by $s$, multiplies to obtain $u$, squares to obtain $t$, evaluates $\theta(t)$ by a piecewise polynomial with Horner's rule, rotates the flag through the bits of $\theta$ by phase-gradient addition, and uncomputes. The angle $\theta(t)$ has a kink at $t=1$, where $g=1$, but is analytic on each side, so the pieces are split at $t=1$. For $t$ beyond a cutoff $t_c$ at which $g<2^{-(b+2)}$ the angle is set to $\pi$. The zero-frequency bin needs no special treatment: $u=0$ gives $t=0$, $g=0$, and $\theta=\pi$, which sends the flag to $\qcwtket1$, as the zero-mean wavelet requires.

The error budget follows from a simple propagation bound. If the implemented entries satisfy $|\widetilde d_{sk}-d_{sk}|\le\delta$ for all $(s,k)$, the unnormalized success branch changes by at most $\delta$ in Euclidean norm, because the error vector has entries $\sqrt{\pi_s}(\widetilde d_{sk}-d_{sk})(Fx)_k/\|x\|_2$ and $\sum_s\pi_s=1$. Hence $|\widetilde P_0^{1/2}-P_0^{1/2}|\le\delta$, and the normalized conditional state, which carries the CWT coefficients, changes by at most $2\delta/P_0^{1/2}$. A small success probability thus amplifies multiplier error, and the precision must be chosen relative to $P_0$.

We emulated the fixed-point pipeline classically for the Doppler example, rounding to $b$ fractional bits after every operation, with scale constants stored to $b+n+1$ bits so that the product with $|\widetilde k|<2^n$ retains $b$ fractional bits. For each $b$ we chose the cheapest number $P$ of polynomial pieces per side of $t=1$ and degree $D$ for which the maximal entry error $\delta$ was below $2^{-(b-2)}$. Table~\ref{tab:arith} reports $\delta$, the observed error of the normalized conditional state, the bound $2\delta/P_0^{1/2}$, the success probability, and a resource estimate. The Toffoli count uses a stated cost model rather than a compiled circuit: $M$ Toffoli gates for the table lookup \citep{Babbush2018}, $n$ for the absolute value, one per bit pair in schoolbook multiplication and squaring, $2P$ per coefficient lookup, $b^2+b$ per Horner step, ripple-carry additions with one Toffoli per bit \citep{Gidney2018}, and $b$ for the phase-gradient rotation, with all computation uncomputed. The work-qubit count includes the frequency, constant, $u$, $t$, coefficient, accumulator, angle, and phase-gradient registers and adder ancillas. With $b=12$, about $3{,}300$ Toffoli gates and 131 work qubits give a conditional-state error of $5\times10^{-4}$, against the $24{,}576$ arbitrary rotations of the table. Under this cost model the total is $M$ plus a term polylogarithmic in $N$ and $M$ for the arithmetic after lookup; for $M=48$ the lookup contributes $48$ of the $3{,}274$ gates, but for dense grids the linear term dominates eventually. On an octave grid with $V$ a power of two, the scale label splits as $s=oV+v$ and $a_s=a_{\min}2^o2^{v/V}$, so a lookup over the $V$ voice constants followed by a binary shift by the octave index $o$ would replace the $M$-entry lookup and make the whole multiplier polylogarithmic; we have not costed that variant. The counts should be taken as order-of-magnitude estimates until the circuit is compiled.

\begin{table}[htbp]
\centering
\caption{Fixed-point emulation of the reversible Mexican-hat multiplier for the 48-scale Doppler example with envelope normalization ($n=9$, $M=48$). $b$, fractional bits; $P$, pieces per side of $t=1$; $D$, polynomial degree; $\delta$, maximal entry error; state error, observed Euclidean error of the normalized conditional state; bound, $2\delta/P_0^{1/2}$. Toffoli gates and work qubits follow the cost model in the text.}
\label{tab:arith}
{\small
\begin{tabular}{rrr|ccc|c|rr}
\toprule
$b$ & $P$ & $D$ & $\delta$ & state error & bound & $P_0$ & Toffoli & qubits\\
\midrule
10 & 1 & 6 & 3.2e-03 & 1.9e-03 & 1.5e-02 & 0.17399 & 2{,}500 & 115 \\
12 & 2 & 6 & 6.4e-04 & 4.8e-04 & 3.1e-03 & 0.17395 & 3{,}274 & 131 \\
14 & 2 & 8 & 1.5e-04 & 1.2e-04 & 7.2e-04 & 0.17395 & 4{,}988 & 147 \\
16 & 4 & 8 & 3.5e-05 & 2.8e-05 & 1.7e-04 & 0.17395 & 6{,}278 & 163 \\
\bottomrule
\end{tabular}
}
\end{table}

\subsection{Amplitude Amplification}\label{sec:aa}

When the success probability is small, amplitude amplification \citep{Brassard2002} boosts it with $O(P_0^{-1/2})$ applications of the full circuit and its inverse, which requires coherent access to $A_x^\dagger$. Because $P_0$ depends on the unknown input, fixed-point amplitude amplification, which needs only a lower bound on $P_0$ and does not overshoot, is the appropriate variant \citep{YoderLowChuang2014}. By Corollary~\ref{cor:log}, an input whose spectrum lies in the covered band has $P_0\approx C_\psi/(\|\widehat\psi\|_\infty^2O\ln2)$ in the independent-prior architecture, so the number of rounds is of order $O^{1/2}$ for the scale-adaptive construction and of order $2^{O/2}$ for the common one. Inputs with substantial energy outside the covered band have smaller $P_0$ in every architecture, by (\ref{eq:globalbound}); removing the sample mean classically before loading costs nothing for a zero-mean wavelet, whose multiplier vanishes at zero frequency, and can raise $P_0$ considerably.

\section{What Can Be Read Out}
\label{sec:readout}

\subsection{The Multinomial Scalogram Model}

A computational-basis measurement of all registers returns the flag and, on success, a pair $(s,j)$. With $n$ successful shots, the counts $n_{sj}$ are multinomial with probabilities (\ref{eq:law}), the number of successes among all shots is binomial with probability (\ref{eq:P0}), and the two are independent given the totals. Let $E_s=\|c_s\|_2^2=\sum_j|c_{sj}|^2$ denote the unweighted energy at scale $s$ and $p_s=w_sE_s/\sum_rw_rE_r$ the scale marginal.

\begin{proposition}\label{prop:multinomial}
The scale counts $n_s=\sum_jn_{sj}$ are multinomial with probabilities $p_s$, and $\hat p_s=n_s/n$ is unbiased with variance $p_s(1-p_s)/n$. For any two scales, the log-contrast $\log(n_s/w_s)-\log(n_r/w_r)$ estimates $\log(E_s/E_r)$ with approximate variance $1/(np_s)+1/(np_r)$. The absolute weighted energy is recovered from the success frequency $\hat P_0$ as $\sum_sw_sE_s=\|x\|_2^2\hat P_0\sum_rw_r\alpha_r^2$.
\end{proposition}

Three practical points follow. First, comparisons intended to estimate a slope of unweighted energy across scales must remove the known weights, as in the log-contrast above; the scale-adaptive normalization does not change this, because the conditional law (\ref{eq:law}) is the same. Second, restricting $j$ to an interval $I$ gives the time-localized energy $\sum_{j\in I}|c_{sj}|^2$, so local scaling and regularity can be estimated from the same record. Third, the scale law is invariant to circular shifts of $x$, because each $c_s$ is a circular convolution; translation invariance is automatic here, whereas in the QNDWT it is produced by tracing out a shift register.

\subsection{A Fourier-Basis Baseline for Global Scale Energies}

Global scale energies do not require the coherent transform. Measuring $F\qcwtket x$ in the computational basis returns $k$ with probability $|(Fx)_k|^2/\|x\|_2^2$, and by (\ref{eq:discretecwt})
$$
\E\,|m_{sk}|^2=\sum_k|m_{sk}|^2\frac{|(Fx)_k|^2}{\|x\|_2^2}=\frac{E_s}{\|x\|_2^2}
$$
for every scale at once. A single frequency outcome therefore yields unbiased estimates of all normalized scale energies, with no flag, no heralding, and no inverse Fourier transform; the multipliers $|m_{sk}|^2$ are evaluated classically. The covariance of the estimates across scales is that of the vector $(|m_{sk}|^2)_s$ under the outcome law, which can be estimated from the same record. This baseline cannot supply time-local information, signs, or phases, which require the translation register of the coherent transform, but it is the natural comparison for global statistics such as scaling exponents. Because the quantum CWT returns a scale sample only on success, the comparison must be made at equal numbers of input preparations, not at equal numbers of successful shots. At that level the comparison can be settled exactly.

\begin{proposition}[Fourier simulation of the scale readout]\label{prop:fourier}
Consider any circuit satisfying the hypotheses of Theorem~\ref{thm:global} with output factor $\gamma$, followed by a measurement of the herald and the scale register. Measure $F\qcwtket x$ in the computational basis and, given the outcome $k$, report success at scale $s$ with probability $|\gamma|^2w_s|m_{sk}|^2$ and failure with the remaining probability $1-|\gamma|^2q_k$. Then the pair (flag, scale) has exactly the law of the circuit's herald and scale outcomes. In particular, for $B$ preparations the QCWT record of flags and scales is a randomization of the Fourier record, and every risk attainable from the former is attainable from the latter \citep{Blackwell1953}.
\end{proposition}

\begin{proof}
The construction is valid because $|\gamma|^2q_k\le|\gamma|^2q_{\max}\le1$ by Theorem~\ref{thm:global}. The synthetic probability of success at $s$ is $\sum_k|\gamma|^2w_s|m_{sk}|^2|(Fx)_k|^2/\|x\|_2^2=|\gamma|^2w_s\|c_s\|_2^2/\|x\|_2^2$ by (\ref{eq:discretecwt}) and unitarity of $F$, which is the probability that the circuit heralds success and returns $s$.
\end{proof}

For the architecture of Fig.~\ref{fig:qcwt-circuit}, $|\gamma|^2=1/\sum_rw_r\alpha_r^2$, and the identity holds for the common, scale-adaptive and envelope normalizations alike; for the two-tone example it reproduces the joint columns of Table~\ref{tab:demo} to within $3\times10^{-17}$. The average of $|m_{sk}|^2$ over the Fourier outcomes, used above, is the conditional expectation of the synthetic success indicator at scale $s$ given $k$, divided by $|\gamma|^2w_s$; it is the Rao--Blackwellized form of the randomized estimator, so for linear functionals of the scale counts it has no larger variance. The proposition concerns the scale marginal only: the translation label of the QCWT has no counterpart in the Fourier record. It also does not order particular nonlinear estimators, such as the fitted Hurst exponents of Section~\ref{sec:hurst}; that comparison remains empirical.

\subsection{A Selected Complex Coefficient}

Define the length-$N$ vector $\kappa_s$ as the ordinary inverse DFT of $\overline{m_{s,\cdot}}$, with factor $1/N$, and let $S^j$ be the circular delay by $j$ samples. Then $c_{sj}=\langle S^j\kappa_s,x\rangle$ with the inner product conjugate-linear in the first argument. Let $A_{sj}$ prepare $S^j\kappa_s/\|\kappa_s\|_2$, assuming $\kappa_s\ne0$. A Hadamard test of $A_{sj}^\dagger A_x$ on $\qcwtket{0^n}$ gives the real and imaginary parts of
$$
\qcwtbra{0^n}A_{sj}^\dagger A_x\qcwtket{0^n}=\frac{c_{sj}}{\|\kappa_s\|_2\|x\|_2},
$$
where the imaginary part is obtained by applying $S^\dagger=\operatorname{diag}(1,-\ii)$ to the test ancilla before its final Hadamard gate \citep{Aharonov2009}. The ancilla returns $0$ with probability $\{1+\Re(\cdot)\}/2$ without the phase gate and $\{1+\Im(\cdot)\}/2$ with it, so the standard error of the real part after $n$ runs is at most $n^{-1/2}$ on the normalized scale. The test does not use the flagged circuit of Fig.~\ref{fig:qcwt-circuit}: it needs a separate circuit $A_{sj}$ preparing the shifted atom, which is classically known, and coherent controlled access to both $A_x$ and $A_{sj}$. It returns $c_{sj}$ relative to $\|\kappa_s\|_2\|x\|_2$; $\|\kappa_s\|_2$ is computed classically, but an unnormalized coefficient, like the absolute energy in Proposition~\ref{prop:multinomial}, also requires the input norm $\|x\|_2$, which the state does not carry. Reading the scalogram gives magnitudes only; signs and phases require interference of this kind.

\subsection{Relation to the Quantum Nondecimated Wavelet Transform}

The QNDWT level shares of \citet{Vidakovic2025} are also quadratic forms in the Fourier coefficients of $x$: the shift-averaged energy of level $j$ equals $\sum_kA_j(\omega_k)|(Fx)_k|^2/\|x\|_2^2$, where $A_j$ is the squared gain of the nondecimated level-$j$ filter normalized so that $\sum_jA_j\equiv1$ together with the scaling band. The QNDWT scale law is therefore the scalogram law of a flagged construction with $w_j=1$ and $|m_{jk}|^2=A_j(\omega_k)$, a Parseval frame with $q_k\equiv1$. Theorem~\ref{thm:optimal} gives such a construction with an independent prior the maximal success probability $1/\sum_j\max_kA_j$, which for the db2 transform with five detail levels and $N=256$ is $0.21$, whereas the bound of Theorem~\ref{thm:global} is one because $q_{\max}=1$. The QNDWT attains that bound without any frequency-conditioned preparation, because the orthogonality of the decimated transforms lets the shift register absorb the redundancy. This is the precise sense in which the two constructions are complementary: orthogonal dyadic filter banks attain deterministic success cheaply, and arbitrary scale grids cost either a postselection whose probability, by Corollary~\ref{cor:log}, decreases as the inverse logarithmic scale range, or a frequency-conditioned scale preparation whose cost is open.

\section{Worked Examples}
\label{sec:examples}

\subsection{A Non-Dyadic Eight-Scale Transform}

We used the Mexican hat with $\widehat\psi(u)=u^2\exp(-u^2/2)$, $N=T=64$ so that $\Delta t=1$, and samples of
$$
x(t)=\cos(2\pi\,3t/T)+0.6\sin(2\pi\,10t/T).
$$
The eight scales are approximately log-spaced, $a_s\in\{0.55,0.80,1.15,1.65,2.40,3.40,4.90,7.00\}$, with weights $w_s\propto1/a_s$. The two tone frequencies have their maximal unweighted CWT magnitudes near $a=\sqrt{5/2}/|\omega|$, that is, at $5.37$ and $1.61$. With the common normalization $\alpha=\max_{s,k}|m_{sk}|=1.943530$ the flag-block evolution gives $P_0=0.064689$; with the scale-adaptive normalization it gives $P_0=0.317128$, a factor of $4.9$, while the scale prior becomes uniform to within $1.3$ percent and the conditional law is unchanged (Table~\ref{tab:demo}). The envelope normalization, with an exactly uniform prior, gives $0.316052$, and the frequency-conditioned bound of Theorem~\ref{thm:global} is $0.986388$. Rescaling the success amplitudes recovers the classical CWT coefficients, including their signs; at $a_s=4.90$ and $b_j=8$ the coefficient is $-1.150396$. The success amplitudes agreed with an FFT-based CWT and with an independent circular atom correlation to within $10^{-15}$ under both normalizations.

\begin{table}[htbp]
\centering
\caption{Ideal scale readout for the two-tone example. The joint columns sum to the flag success probability for common and scale-adaptive normalization; the conditional column is the weighted scale-energy distribution, identical under both.}
\label{tab:demo}
\begin{tabular}{rrrr}
\toprule
$a_s$ & $\Pr(s,0)$, common & $\Pr(s,0)$, adaptive & $\Pr(s\mid 0)$\\
\midrule
0.55 & 0.000818 & 0.004009 & 0.012642\\
0.80 & 0.002670 & 0.013090 & 0.041277\\
1.15 & 0.006090 & 0.029853 & 0.094136\\
1.65 & 0.007777 & 0.038124 & 0.120215\\
2.40 & 0.006760 & 0.033142 & 0.104505\\
3.40 & 0.012842 & 0.062957 & 0.198521\\
4.90 & 0.018781 & 0.092071 & 0.290328\\
7.00 & 0.008951 & 0.043882 & 0.138375\\
\midrule
Total & 0.064689 & 0.317128 & 1.000000\\
\bottomrule
\end{tabular}
\end{table}

\subsection{A 48-Scale Analysis of the Doppler Signal}\label{sec:doppler}

\begin{figure}[htbp]
\centering
\includegraphics[width=\linewidth]{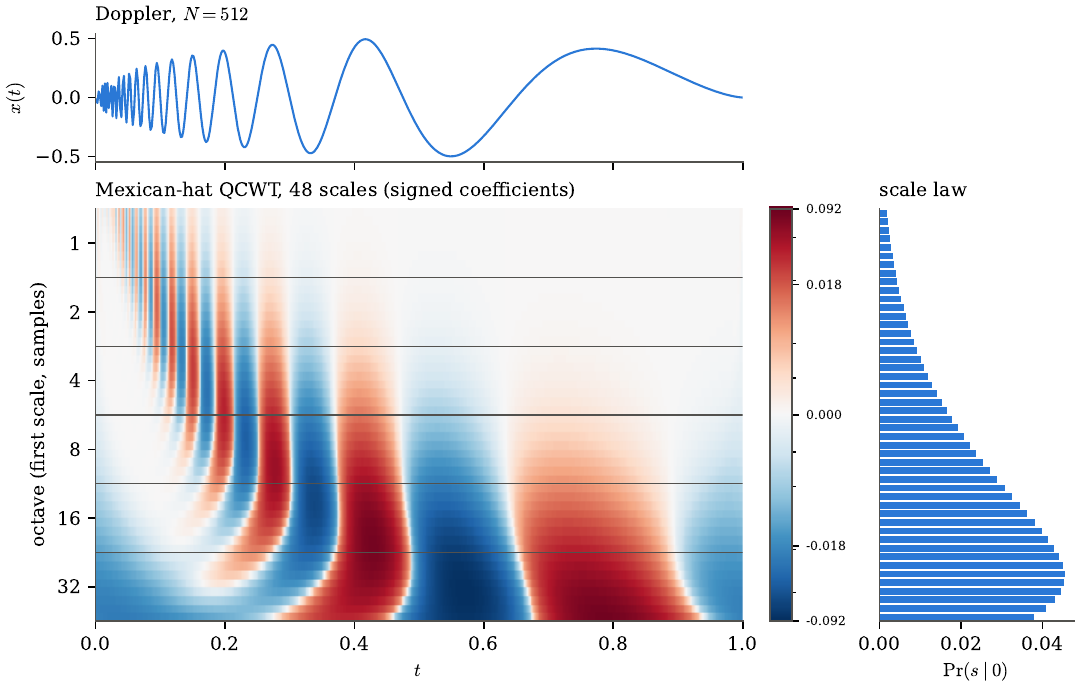}
\caption{Doppler signal, $N=512$ (top); signed Mexican-hat QCWT coefficients reconstructed from the simulated success amplitudes after removing the known normalization, which a measurement does not return for 48 scales, six octaves of eight voices beginning at 1, 2, 4, 8, 16, and 32 sample spacings, on a signed nonlinear color scale (bottom left); conditional scale law $\Pr(s\mid0)$, which is what computational-basis measurements of the success branch sample, after summation over translations (bottom right).}
\label{fig:doppler}
\end{figure}

The Doppler signal of \citet{DonohoJohnstone1994} was sampled at $N=512$ cell centers and analyzed with 48 Mexican-hat scales, six octaves of eight voices starting at one sample spacing, with weights $w_s\propto1/a_s$ and all 512 translations per scale. The explicit simulation of the scale, data, and flag amplitudes reproduced the direct circular correlation to within $1.3\times10^{-16}$ (Fig.~\ref{fig:doppler}). The common normalization gives $P_0=0.0120$ and a scale prior ranging from $0.0014$ to $0.084$; the scale-adaptive normalization gives $P_0=0.176$, a factor of $14.6$, with a prior between $0.0186$ and $0.0210$; the envelope normalization with a uniform prior gives $0.174$; and the frequency-conditioned bound is $0.783$. The independent-prior value is close to $\rho\approx1.33/O=0.22$ of Corollary~\ref{cor:log} for $O=6$, which the Doppler signal does not attain because part of its energy lies at frequencies covered by fewer scales near the ends of the grid. The right panel of Fig.~\ref{fig:doppler} shows the conditional scale law, which concentrates on the coarse octaves where the slow oscillations of the Doppler signal carry most of its energy.

\subsection{Complex-Valued Multipliers and Coefficients}
\label{sec:complex}

Both preceding examples use the Mexican hat, whose multiplier is real and nonnegative and whose coefficients are real. Two further wavelets test the phase-sensitive parts of the construction. The first derivative of a Gaussian, $\psi(t)=-te^{-t^2/2}$, has $\widehat\psi(u)=\ii\sqrt{2\pi}\,ue^{-u^2/2}$ (we drop the constant), so its multiplier $m_{sk}=\sqrt{a_s}\,\overline{\widehat\psi(a_s\omega_k)}$ is purely imaginary, the conjugation in (\ref{eq:discretecwt}) changes its sign, and the phase gates of Remark~1 are required with $\phi=\pm\pi/2$; its coefficients are real because the multiplier is odd and conjugate-symmetric. The analytic Morlet wavelet, $\widehat\psi(u)=e^{-(u-6)^2/2}$ for $u>0$ and zero otherwise, has a real multiplier but complex coefficients. Because $\widehat\psi(0^+)=e^{-18}\ne0$, its admissibility integral $C_+$ diverges logarithmically at zero frequency, so strictly it is a one-sided Morlet multiplier on the sampled grid rather than an admissible mother wavelet. The sampled multiplier is well defined, since the zero-frequency bin is zero by definition and the smallest nonzero frequency is $2\pi/T$, and all finite-grid quantities reported below are exact; the approximation of Corollary~\ref{cor:log} applies only to admissible mothers. For the two-tone signal and the eight scales of Table~\ref{tab:demo}, the flagged evolution reproduced the direct circular correlation to within $7\times10^{-16}$ for both wavelets, with $P_0=0.4520$ for the derivative of a Gaussian and $0.0219$ for the Morlet wavelet, whose center frequency places most of these scales away from the two tones.

Table~\ref{tab:hadamard} verifies the Hadamard-test readout of Section~\ref{sec:readout}. For each selected coefficient we constructed unitaries $A_x$ and $A_{sj}$ with the required first columns, simulated the test with and without the phase gate $S^\dagger$ on the ancilla, and recovered $c_{sj}$ from the two ancilla probabilities. The recovered values agree with the direct correlation to within $3\times10^{-15}$, including a Morlet coefficient with real and imaginary parts of equal magnitude and opposite sign. With $n$ runs per quadrature, the standard error of each part on the scale of $c_{sj}$ is at most $\|\kappa_s\|_2\|x\|_2n^{-1/2}$.

\begin{table}[htbp]
\centering
\caption{Hadamard-test readout of selected complex coefficients for the two-tone signal. $p_{\Re}$ and $p_{\Im}$ are the ideal probabilities of ancilla outcome $0$ without and with the phase gate $S^\dagger$; error is the absolute difference between the coefficient recovered from them and the direct circular correlation.}
\label{tab:hadamard}
{\small
\begin{tabular}{lcc|cc|cc|c}
\toprule
Wavelet & $a_s$ & $j$ & $\Re c_{sj}$ & $\Im c_{sj}$ & $p_{\Re}$ & $p_{\Im}$ & error\\
\midrule
Morlet & 4.90 & 8 & $+0.3273$ & $+0.0000$ & 0.5467 & 0.5000 & 2e-15 \\
Morlet & 4.90 & 20 & $+0.2315$ & $-0.2315$ & 0.5330 & 0.4670 & 7e-16 \\
DOG1 & 1.65 & 8 & $+0.3922$ & $-0.0000$ & 0.5792 & 0.5000 & 1e-15 \\
DOG1 & 4.90 & 20 & $-0.4316$ & $-0.0000$ & 0.4129 & 0.5000 & 9e-16 \\
\bottomrule
\end{tabular}
}
\end{table}

\begin{figure}[htbp]
\centering
\includegraphics[width=\linewidth]{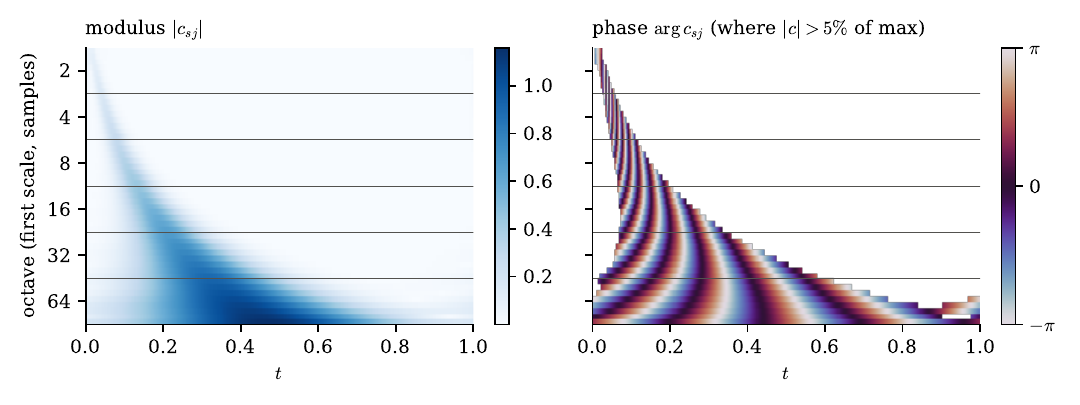}
\caption{Analytic Morlet QCWT of the Doppler signal, $N=512$, 48 scales from 2 to 118 sample spacings, reconstructed from simulated success amplitudes: modulus (left) and phase where the modulus exceeds five percent of its maximum (right).}
\label{fig:morlet}
\end{figure}

Figure~\ref{fig:morlet} shows the analytic Morlet transform of the Doppler signal on 48 scales from 2 to 118 sample spacings, reconstructed from simulated success amplitudes. The modulus concentrates along the time-varying frequency of the signal, and the phase advances along each scale at the local frequency. A measurement of the success branch samples only the squared modulus; the phase panel is available only through interference, for example by Hadamard tests of selected coefficients. The analytic wavelet discards the negative-frequency half of a real signal, so for this example the independent-prior optimum is $P_0=0.016$ and the frequency-conditioned bound is $0.221$, both below one half.

\subsection{What the Joint Law Adds: A Chirp Pair}
\label{sec:chirp}

Proposition~\ref{prop:fourier} shows that the flag and scale outcomes carry nothing beyond a Fourier-basis measurement. The translation label is different, and a measurement of $\qcwtket x$ in the time basis does not replace it either. Let $N=512$, $\tau=t-N/2$, $g_\tau=\exp\{-\tau^2/(2\cdot80^2)\}$, and
$$
x^{\pm}_t=g_\tau\exp\{\ii(\omega_0\tau\pm\beta\tau^2)\},\qquad \omega_0=2\pi\cdot64/N,\quad \beta=2\pi\cdot0.1/512,
$$
an up-chirp and a down-chirp whose instantaneous frequency sweeps between about $0.05$ and $0.20$ cycles per sample over the support of the envelope. Because $\omega_0N/(2\pi)$ is an integer, $x^-_t=\overline{x^+_{-t\bmod N}}$ exactly, so $(Fx^-)_k=\overline{(Fx^+)_k}$ and the two Fourier-basis laws coincide; because $g$ is symmetric, $|x^-_t|=|x^+_t|$ and the two time-basis laws coincide. Numerically both agree to within $3\times10^{-17}$, and so do the QCWT scale marginals, as Proposition~\ref{prop:fourier} requires.

With the 48-scale Mexican-hat grid of Section~\ref{sec:doppler} and the envelope normalization, both inputs have $P_0=0.1953$, but the joint scale--translation laws are at total-variation distance $0.269$. The difference is in where each scale's energy occurs. On the finest octave, $1\le a_s<2$ samples, which carries $59\%$ of the conditional law, the fraction of that octave's mass in the first half of the record is $0.408$ for the up-chirp and $0.585$ for the down-chirp; on the next octave, which carries $40\%$, it is $0.684$ and $0.309$. Summed over scales the two halves are nearly balanced, $0.525$ and $0.468$, so it is the conditional time distribution at fixed scale, not a time marginal, that separates the inputs. The Bhattacharyya coefficient of the two joint laws is $0.900$, so the likelihood-ratio test of $x^+$ against $x^-$ errs with probability at most $\tfrac12(0.900)^n$ after $n$ successful shots, below $0.003$ at $n=50$, or about $250$ preparations at $P_0=0.195$; no number of Fourier-basis or time-basis measurements can perform this test.

\section{Hurst Exponent Estimation with Dense Scale Grids}
\label{sec:hurst}

For fractional Brownian motion $B_H$ with variance parameter $\sigma^2$ and a real wavelet with at least one vanishing moment and sufficient decay, the nonstationary terms of the covariance cancel and
$$
\E|W_{B_H}(a,b)|^2=\sigma^2K_{\psi,H}a^{2H+1},\qquad K_{\psi,H}=-\frac12\int_{\mathbb R}\int_{\mathbb R}\psi(u)\psi(v)|u-v|^{2H}\,du\,dv,
$$
so the unweighted scale energy $E_s$ grows like $a_s^{2H+1}$ and the slope of $\log_2E_s$ against $\log_2a_s$ identifies $H$ \citep{Flandrin1992,AbryVeitch1998,VeitchAbry1999,Bardet2002}. On a finite grid we use the exact moments of the energies rather than this asymptotic form. If $x$ is the vector of path values and $B_s$ maps it, through the mirror extension and the periodic CWT, to the coefficients at scale $s$, then $E_s=x^\top A_sx$ with $A_s=\operatorname{Re}(B_s^\dagger B_s)$, and for a Gaussian path with covariance $\Sigma_H$
\begin{eqnarray}
e_s(H)=\E E_s=\operatorname{tr}(A_s\Sigma_H),\qquad \Gamma_{sr}(H)=\operatorname{Cov}(E_s,E_r)=2\operatorname{tr}(A_s\Sigma_HA_r\Sigma_H).
\label{eq:moments}
\end{eqnarray}
These moments of the energies are exact and absorb finite-sample, boundary, and discretization effects. The moments of their logarithms are not exact: by the delta method, $z_s=\log E_s$ has mean approximately $\log e_s(H)$, up to an additive constant $\log\sigma^2$, and covariance approximately $V_{sr}(H)=\Gamma_{sr}(H)/\{e_s(H)e_r(H)\}$. We compare four estimators that use exact energies. Ordinary least squares regresses $\log_2E_s$ on $\log_2a_s$ and sets $\hat H=(\text{slope}-1)/2$. Calibrated ordinary least squares minimizes $\|z-\log e(H)-c\mathbf 1\|^2$ over $H$ and $c$, which corrects the mean curve but ignores covariance. Calibrated generalized least squares minimizes $\{z-\log e(H)-c\mathbf 1\}^\top V(H)^{-1}\{z-\log e(H)-c\mathbf 1\}$, and a regularized version floors the eigenvalues of $V(H)$ at one percent of the largest. Comparing the first two isolates the effect of the mean curve, and comparing the last two with the second isolates the effect of covariance weighting.

Adding voices costs few qubits, since the scale register grows only as $\log_2M$, but synthesis accuracy, shots per scale, and statistical conditioning still matter. Adjacent Mexican-hat scales with ratio $2^{1/V}$ have nearly collinear filters, and the condition number of $V(H)$ for four octaves grows from about 20 to 30 at $V=1$ to $3\times10^3$ to $9\times10^3$ at $V=2$, $4\times10^8$ to $2\times10^9$ at $V=4$, and numerical singularity at $V=8$. Varying the eigenvalue floor between $10^{-4}$ and $3\times10^{-2}$ changed the standard deviation of $\hat H$ by at most $0.014$ in a pilot run.

For the quantum comparisons we fix the number $B$ of input preparations. The quantum CWT uses the envelope normalization with a uniform prior; each preparation succeeds with probability $P_0$, and the successful shots are multinomial over scales with probabilities $p_s\propto w_sE_s$, from which $z_s=\log(n_s/w_s)$ and the multinomial covariance $\operatorname{diag}\{1/(n_{\rm succ}p_s)\}$ is added to $V(H)$. The Fourier baseline of Section~\ref{sec:readout} uses all $B$ preparations, estimates $E_s/\|x\|_2^2$ by the average of $|m_{sk}|^2$ over the outcomes, and adds the estimated covariance of the logarithms of these averages to $V(H)$. Both use the regularized covariance. A scale with no counts is assigned one half count; no preparation budget produced zero successes.

The simulation used paths of length $N=512$ generated by circulant embedding \citep{DaviesHarte1987} with $H\in\{0.3,0.7\}$, centered, mirror-extended to length 1024 (ten data qubits), and analyzed with the Mexican hat on four octaves of scales $a_s=2\cdot2^{s/V}$ sample spacings, $s=0,\ldots,4V-1$, for $V\in\{1,2,4,8\}$, that is, $M\in\{4,8,16,32\}$ scales and two to five scale qubits. Centering leaves every scale energy unchanged, because the multiplier vanishes at zero frequency, and raises the success probability. The moments (\ref{eq:moments}) were computed exactly on a grid of $H$ with step $0.05$ and interpolated linearly, and each criterion was minimized over a grid with step $0.005$. Table~\ref{tab:hurst} reports root mean squared errors over 600 replications with random seed 2026; the largest Monte Carlo standard error of an entry, obtained by the delta method from the replicate squared errors, is 0.0073. The code is included with the paper.

\begin{table}[htbp]
\centering
\caption{Root mean squared error of $\hat H$ for centered fractional Brownian paths of length 512, Mexican-hat QCWT with mirror extension, four octaves with $V$ voices ($M$ scales), 600 replications. Exact energies: OLS, ordinary least squares; cal.\ OLS, least squares with the exact mean curve; GLS, calibrated generalized least squares; floor, GLS with eigenvalues floored at one percent of the largest. Equal preparations: QCWT, scale counts from the successful shots among $B$ preparations; QFT, Fourier-basis baseline with $B$ preparations; both with GLS floor.}
\label{tab:hurst}
{\small\setlength{\tabcolsep}{4pt}
\begin{tabular}{ccc|cccc|cc|cc}
\toprule
 & & & \multicolumn{4}{c|}{Exact energies} & \multicolumn{2}{c|}{$B=10^4$} & \multicolumn{2}{c}{$B=10^5$}\\
$H$ & $V$ & $M$ & OLS & cal.\ OLS & GLS & floor & QCWT & QFT & QCWT & QFT\\
\midrule
0.3 & 1 & 4 & 0.097 & 0.073 & 0.061 & 0.061 & 0.071 & 0.062 & 0.062 & 0.061 \\
 & 2 & 8 & 0.100 & 0.077 & 0.054 & 0.056 & 0.072 & 0.058 & 0.060 & 0.056 \\
 & 4 & 16 & 0.102 & 0.080 & 0.073 & 0.057 & 0.074 & 0.059 & 0.062 & 0.057 \\
 & 8 & 32 & 0.104 & 0.081 & 0.408 & 0.059 & 0.078 & 0.060 & 0.063 & 0.059 \\
\midrule
0.7 & 1 & 4 & 0.093 & 0.086 & 0.071 & 0.071 & 0.131 & 0.097 & 0.090 & 0.075 \\
 & 2 & 8 & 0.097 & 0.089 & 0.066 & 0.070 & 0.121 & 0.091 & 0.093 & 0.075 \\
 & 4 & 16 & 0.100 & 0.091 & 0.099 & 0.071 & 0.149 & 0.091 & 0.113 & 0.074 \\
 & 8 & 32 & 0.102 & 0.092 & 0.376 & 0.072 & 0.207 & 0.093 & 0.173 & 0.074 \\
\bottomrule
\end{tabular}
}
\end{table}

Five conclusions emerge. First, both the mean curve and the covariance matter, in proportions that depend on $H$. Calibrating the mean curve alone reduces the error of ordinary least squares and removes most of its bias at $H=0.3$, from $-0.03$ to $-0.04$ to about $-0.005$; it accounts for 56 to 73 percent of the reduction in mean squared error achieved by regularized generalized least squares at $H=0.3$, and for about 35 percent at $H=0.7$, where covariance weighting contributes the larger part. Paired over replications, the reduction in squared error from the mean curve is between -0.0042 and -0.0013 with standard errors at most 0.0004, and the further reduction from covariance weighting is between -0.0034 and -0.0016 with standard errors at most 0.0004. Second, the best unregularized result is obtained at $V=2$; at $V=4$ and $V=8$ unregularized generalized least squares degrades, a direct consequence of the near singularity of the covariance. Third, once the covariance is regularized the error is essentially flat in $V$: additional voices add almost no information about a global scaling exponent, because the energies of nearby scales are nearly linear functions of one another. Fourth, at equal numbers of input preparations the Fourier baseline is at least as accurate as the quantum CWT in every setting, as Proposition~\ref{prop:fourier} leads one to expect: the scale counts of the QCWT are a randomization of the Fourier record. The success probability of the quantum CWT averages 0.076 to 0.090 at $H=0.3$ and only 0.017 to 0.026 at $H=0.7$, because a path with large $H$ carries most of its energy at frequencies below the covered band, and the dense grids then leave scales with no counts (4345 zero scale counts in total over the two budgets at $H=0.7$, $V=8$, against none at $H=0.3$ for $V\le4$). At $B=10^5$ and $H=0.3$ both quantum readouts are within $0.005$ of the exact-energy error, but at $H=0.7$ only the Fourier baseline is. Fifth, therefore, the quantum CWT earns its cost for time-local and phase-sensitive quantities, not for global scale energies.

The practical recommendation is one or two voices per octave for estimating a global Hurst exponent, calibrated generalized least squares with exact energy moments and a regularized covariance, and, on a quantum device, the Fourier-basis measurement for this global statistic. Dense grids and the coherent transform are better reserved for questions that need fine scale and time resolution, such as magnitude ridges, narrow-band energy comparisons, local regularity, and phase.

\section{Resources, Scope, and Open Problems}
\label{sec:discussion}

The logical construction uses $n$ data qubits, $\lceil\log_2M\rceil$ scale qubits, one flag, and two quantum Fourier transforms with $O(n^2)$ gates \citep{coppersmith1994,nielsen2010}; a mirror extension adds one qubit and $n$ controlled NOT gates. With the envelope normalization on a uniform logarithmic grid the scale preparation is a uniform superposition. The multiplier gate costs $MN$ rotations as a table and, by the estimate of Table~\ref{tab:arith}, a few thousand Toffoli gates and about 130 work qubits for a conditional-state error of order $10^{-3}$ to $10^{-4}$ in the Doppler example by reversible arithmetic. Under the stated cost model that estimate contains an $M$-entry lookup, so it is linear in $M$ plus polylogarithmic arithmetic; the octave--voice factorization of Section~\ref{sec:synth} would remove the linear term, but it has not been costed. The success probability of the independent-prior architecture is the global bound (\ref{eq:globalbound}) times $\rho\approx C_\psi/(\|\widehat\psi\|_\infty^2O\ln2)$, so the expected number of preparations per successful shot grows linearly in the number of octaves for in-band inputs, and quadratically more slowly with amplitude amplification; a frequency-conditioned scale preparation would remove the factor $\rho$ at a cost that remains to be determined. These costs exclude the preparation of $\qcwtket x$, which is $\Theta(N)$ for arbitrary classical data, so end-to-end advantages require a coherent data source, and full classical recovery of the $MN$ coefficients forfeits any advantage.

Within these limits, the comparison with the Fourier baseline locates the contribution. Global scale energies, and statistics built from them such as scaling exponents, are estimated at least as well from Fourier-basis measurements of the input, without heralding; by Proposition~\ref{prop:fourier} this holds at the level of the measurement model, for every heralded construction covered by Theorem~\ref{thm:global}. The coherent transform contributes what that baseline cannot: the joint scale--translation law, which carries time-local energy and supports magnitude-based ridge and local-regularity analysis directly from measurement samples, as the chirp pair of Section~\ref{sec:chirp} illustrates, and the complex coefficients themselves, whose phases are accessible by interference. The scale register grows only logarithmically in the number of scales, but dense grids are not free statistically, as Section~\ref{sec:hurst} shows.

Several problems remain open. One is a compiled gate-level count of the arithmetic multiplier for standard wavelets, which would replace the cost model of Table~\ref{tab:arith}, and an efficient realization, or a lower bound, for the frequency-conditioned scale preparation of Theorem~\ref{thm:global}. Another concerns ridges. Ridges defined by the scalogram magnitude can be estimated from samples of the joint law (\ref{eq:law}), whereas phase-based estimates of instantaneous frequency require the phases of analytic-wavelet coefficients and therefore interference measurements such as those of Table~\ref{tab:hadamard}, at a cost per coefficient. A third is the extension to two-dimensional and directional continuous transforms, where the scale register becomes a register of scales and orientations and the frame identity (\ref{eq:frame}) again governs the success probability.

\paragraph{Acknowledgments.}
 Supported by the National Science Foundation under Grant No.~2515246 and by the H.~O.~Hartley Chair funds at Texas A\&M University.

\appendix
\section{Reproducible NumPy Verification}

The code evolves the scale, frequency, and flag amplitudes explicitly for both normalizations and checks the result against an independent circular atom correlation. The scripts that produce every table and figure, including the fixed-point emulation, the Hurst study with seed 2026, and the chirp pair of Section~\ref{sec:chirp} (\texttt{chirp\_pair.py}), are supplied 
via \url{https://github.org/BraniV/QuantumCWT.}

\begin{lstlisting}
import numpy as np
N, T = 64, 64.0
t = T*np.arange(N)/N
omega = 2*np.pi*np.fft.fftfreq(N, d=T/N)
a = np.array([0.55, 0.80, 1.15, 1.65, 2.40, 3.40, 4.90, 7.00])
f = np.cos(3*2*np.pi*t/T) + .6*np.sin(10*2*np.pi*t/T)
m = np.sqrt(a[:,None])*(a[:,None]*omega)**2*np.exp(-.5*(a[:,None]*omega)**2)
w = np.log(1.45)/a
lag = (np.arange(N)[None,:] - np.arange(N)[:,None]) % N
direct = np.stack([atom[lag].conj() @ f for atom in np.fft.ifft(m.conj(), axis=1)])
for alpha in [np.full(a.size, abs(m).max()), abs(m).max(1)]:   # common, adaptive
    pi = w*alpha**2; pi /= pi.sum()
    d = m/alpha[:,None]; r = np.sqrt(np.clip(1-abs(d)**2, 0, 1))
    B = np.empty(m.shape+(2,2), complex)
    B[...,0,0], B[...,0,1], B[...,1,0], B[...,1,1] = d, -r, r, d.conj()
    st = np.zeros(m.shape+(2,), complex)
    st[...,0] = np.sqrt(pi[:,None])*np.fft.fft(f/np.linalg.norm(f), norm="ortho")
    st = np.fft.ifft(np.einsum("skij,skj->ski", B, st), axis=1, norm="ortho")
    succ = st[...,0]; P0 = np.sum(abs(succ)**2)
    coef = succ*alpha[:,None]*np.linalg.norm(f)/np.sqrt(pi[:,None])
    np.testing.assert_allclose(coef, direct, atol=1e-13)
    print(f"P0 = {P0:.6f}", np.round(np.sum(abs(succ)**2, 1), 6))
\end{lstlisting}


\begin{thebibliography}{99}
\bibitem[Aaronson(2015)]{Aaronson2015}
Aaronson, S. (2015). Read the fine print. \emph{Nature Physics}, 11, 291--293.

\bibitem[Abry and Veitch(1998)]{AbryVeitch1998}
Abry, P. and Veitch, D. (1998). Wavelet analysis of long-range-dependent traffic. \emph{IEEE Transactions on Information Theory}, 44, 2--15.

\bibitem[Aharonov et~al.(2009)]{Aharonov2009}
Aharonov, D., Jones, V., and Landau, Z. (2009). A polynomial quantum algorithm for approximating the Jones polynomial. \emph{Algorithmica}, 55, 395--421.

\bibitem[Babbush et~al.(2018)]{Babbush2018}
Babbush, R., Gidney, C., Berry, D.~W., Wiebe, N., McClean, J., Paler, A., Fowler, A., and Neven, H. (2018). Encoding electronic spectra in quantum circuits with linear T complexity. \emph{Physical Review X}, 8, 041015.

\bibitem[Bagherimehrab and Aspuru-Guzik(2024)]{BagherimehrabAspuruGuzik2024}
Bagherimehrab, M. and Aspuru-Guzik, A. (2024). Efficient quantum algorithm for all quantum wavelet transforms. \emph{Quantum Science and Technology}, 9. \href{https://doi.org/10.1088/2058-9565/ad3d7f}{doi:10.1088/2058-9565/ad3d7f}.

\bibitem[Bardet(2002)]{Bardet2002}
Bardet, J.-M. (2002). Statistical study of the wavelet analysis of fractional Brownian motion. \emph{IEEE Transactions on Information Theory}, 48, 991--999.

\bibitem[Blackwell(1953)]{Blackwell1953}
Blackwell, D. (1953). Equivalent comparisons of experiments. \emph{Annals of Mathematical Statistics}, 24, 265--272.

\bibitem[Brassard et~al.(2002)]{Brassard2002}
Brassard, G., H{\o}yer, P., Mosca, M., and Tapp, A. (2002). Quantum amplitude amplification and estimation. \emph{Contemporary Mathematics}, 305, 53--74. \href{https://arxiv.org/abs/quant-ph/0005055}{arXiv:quant-ph/0005055}.

\bibitem[Childs and Wiebe(2012)]{ChildsWiebe2012}
Childs, A.~M. and Wiebe, N. (2012). Hamiltonian simulation using linear combinations of unitary operations. \emph{Quantum Information and Computation}, 12, 901--924.

\bibitem[Coppersmith(1994)]{coppersmith1994}
Coppersmith, D. (1994). An approximate Fourier transform useful in quantum factoring. IBM Research Report RC19642. \href{https://arxiv.org/abs/quant-ph/0201067}{arXiv:quant-ph/0201067}.

\bibitem[Daubechies(1992)]{Daubechies1992}
Daubechies, I. (1992). \emph{Ten Lectures on Wavelets}. SIAM.

\bibitem[Davies and Harte(1987)]{DaviesHarte1987}
Davies, R.~B. and Harte, D.~S. (1987). Tests for Hurst effect. \emph{Biometrika}, 74, 95--101.

\bibitem[Donoho and Johnstone(1994)]{DonohoJohnstone1994}
Donoho, D.~L. and Johnstone, I.~M. (1994). Ideal spatial adaptation by wavelet shrinkage. \emph{Biometrika}, 81, 425--455.

\bibitem[Flandrin(1992)]{Flandrin1992}
Flandrin, P. (1992). Wavelet analysis and synthesis of fractional Brownian motion. \emph{IEEE Transactions on Information Theory}, 38, 910--917.

\bibitem[Gidney(2018)]{Gidney2018}
Gidney, C. (2018). Halving the cost of quantum addition. \emph{Quantum}, 2, 74.

\bibitem[Gily\'en et~al.(2019)]{Gilyen2019}
Gily\'en, A., Su, Y., Low, G.~H., and Wiebe, N. (2019). Quantum singular value transformation and beyond: Exponential improvements for quantum matrix arithmetics. \emph{Proceedings of the 51st Annual ACM SIGACT Symposium on Theory of Computing}, 193--204. \href{https://arxiv.org/abs/1806.01838}{arXiv:1806.01838}.

\bibitem[Grossmann and Morlet(1984)]{GrossmannMorlet1984}
Grossmann, A. and Morlet, J. (1984). Decomposition of Hardy functions into square integrable wavelets of constant shape. \emph{SIAM Journal on Mathematical Analysis}, 15, 723--736.

\bibitem[Grover and Rudolph(2002)]{GroverRudolph2002}
Grover, L. and Rudolph, T. (2002). Creating superpositions that correspond to efficiently integrable probability distributions. \href{https://arxiv.org/abs/quant-ph/0208112}{arXiv:quant-ph/0208112}.

\bibitem[Mallat(1999)]{Mallat1999}
Mallat, S. (1999). \emph{A Wavelet Tour of Signal Processing}, 2nd ed. Academic Press.

\bibitem[M{\"o}tt{\"o}nen et~al.(2004)]{Mottonen2004}
M{\"o}tt{\"o}nen, M., Vartiainen, J.~J., Bergholm, V., and Salomaa, M.~M. (2004). Quantum circuits for general multiqubit gates. \emph{Physical Review Letters}, 93, 130502.

\bibitem[Nielsen and Chuang(2010)]{nielsen2010}
Nielsen, M.~A. and Chuang, I.~L. (2010). \emph{Quantum Computation and Quantum Information}. Cambridge University Press.

\bibitem[Plesch and Brukner(2011)]{PleschBrukner2011}
Plesch, M. and Brukner, {\v C}. (2011). Quantum-state preparation with universal gate decompositions. \emph{Physical Review A}, 83, 032302.

\bibitem[Shukla and Vedula(2024)]{ShuklaVedula2024}
Shukla, A. and Vedula, P. (2024). An efficient quantum algorithm for preparation of uniform quantum superposition states. \href{https://arxiv.org/abs/2306.11747}{arXiv:2306.11747}.

\bibitem[Torrence and Compo(1998)]{TorrenceCompo1998}
Torrence, C. and Compo, G.~P. (1998). A practical guide to wavelet analysis. \emph{Bulletin of the American Meteorological Society}, 79, 61--78.

\bibitem[Veitch and Abry(1999)]{VeitchAbry1999}
Veitch, D. and Abry, P. (1999). A wavelet-based joint estimator of the parameters of long-range dependence. \emph{IEEE Transactions on Information Theory}, 45, 878--897.

\bibitem[Vidakovic(2025)]{Vidakovic2025}
Vidakovic, B. (2025). Quantum nondecimated wavelet transform: Theory, circuits, and applications. \href{https://arxiv.org/abs/2512.21478}{arXiv:2512.21478}.

\bibitem[Yoder et~al.(2014)]{YoderLowChuang2014}
Yoder, T.~J., Low, G.~H., and Chuang, I.~L. (2014). Fixed-point quantum search with an optimal number of queries. \emph{Physical Review Letters}, 113, 210501.

\bibitem[Zhang et~al.(2022)]{ZhangLiYuan2022}
Zhang, X.-M., Li, T., and Yuan, X. (2022). Quantum state preparation with optimal circuit depth: Implementations and applications. \emph{Physical Review Letters}, 129, 230504.

\end{thebibliography}
\end{document}